\documentclass[aps,pre,twocolumn,groupedaddress]{revtex4-2}
\usepackage{bm}
\usepackage{amsmath}
\usepackage{graphicx}
\usepackage{hyperref}
\usepackage{xcolor, soul}
\begin{document}

\title{Rayleigh Taylor Instability in Multiple Finite-Thickness Fluid Layers}
\author{Prashant Sharma}
\email[]{sharmap@bu.edu}
%\thanks{}
\altaffiliation{Department of Physics, Suffolk University, 8 Ashburton Pl., Boston, MA 02108}
\affiliation{Department of Physics, Boston University 590 Commonwealth Avenue, Boston, MA 02215}
\date{\today}
\begin{abstract}
We develop a general transfer-matrix formalism for determining the growth rate of the Rayleigh-Taylor instability in a fluid system with spatially varying density and viscosity. We use this formalism to analytically and numerically treat the case of a stratified heterogeneous fluid. We introduce the inviscid-flow approximation in our transfer-matrix formalism to find analytic solutions in the limit of uniform kinematic viscosity for a stratified heterogeneous fluid. We discuss the applicability of these results and a related approximation that also yields analytical solutions in the large viscosity limit. 
\end{abstract}
\maketitle
 The Rayleigh-Taylor instability is a hydrodynamic instability of the interface of a dense fluid supported by a lower-density fluid in gravity~\cite{Rayleigh_Investigation_1882} or the presence of upward acceleration~\cite{Taylor_Instability_1950}. The instability occurs in many natural processes that arise in gas-particle mixtures~\cite{Balakrishnan_Explosion_2014}, radiation pressure acceleration of plasma~\cite{Wilks_Absorption_1992,Gamaly_Instability_1993,Pegoraro_Photon_2007,Palmer_Rayleigh-Taylor_2012}, astrophysical structures like supernova remnants~\cite{Chevalier_Outer_1975, Ribeyre_Compressible_2004,Joggerst_Three-dimensional_2010, Porth_Rayleigh-Taylor_2014}, geophysical processes~\cite{Plag_Rayleigh-Taylor_1995, Conrad_Growth_1997,Lev_Rayleigh-Taylor_2008,Mondal_Propagator_2017}, and capsule implosions in inertial confinement fusion~\cite{Nuckolls_Laser_1972,Lindl_Development_1995,Hurricane_Fuel_2014,Campbell_Laser_2017}.
 
  The classical theory for two fluids with a sharp interface~\cite{Chandrasekhar_Hydrodynamic_1961} is focused primarily on semi-infinite fluids. More recent works~\cite{Mikaelian_Effect_1993,Mikaelian_Rayleigh-Taylor_1996,Piriz_Rayleigh-Taylor_2006} have introduced the inviscid-flow approximation in the classical linear perturbation theory that yields analytic and physically insightful~\cite{Piriz_Rayleigh-Taylor_2006} solutions for the two-fluid case with finite layer thickness. 
  
  In this paper, we generalize the classical theory for two fluids to a fluid system with spatially varying density and viscosity. We treat this system as a stratified heterogeneous fluid, each layer having a constant density. This mapping allows us to study the temporal growth of the interfacial perturbations in a transfer-matrix formalism that incorporates the boundary conditions of the fluid system. While this formalism is exact, it only yields a computational strategy. We use the inviscid-flow approximation in the transfer-matrix formalism and show that in the case of uniform viscosity, or zero-viscosity of the fluid system with $M$-interfaces, the problem of finding the growth rate reduces to finding the eigenvalues of an $M\times M$ matrix that only depends on the unperturbed fluid properties. The simplified formulation applies to problems with spatially varying density. Such systems comprise most of the natural processes in which the Rayleigh-Taylor instability occurs. Thus, for example, in ablative Rayleigh-Taylor instability in inertial confinement fusion process~\cite{Betti_Growth_1998}, there is no discontinuous density jump, and the classical theory does not apply. Using this formalism, we recover the growth rate dispersion for a fluid system with smoothly varying density: $\gamma=\sqrt{A_T kg/(1+A_T k L_m)}$, where $A_T$ is the Atwood ratio (of density difference to the density sum), and $L_m$ is the density gradient length scale~\cite{Bud'ko_Stabilization_1992, Betti_Growth_1998, Yu_Multiple_2018, Dong_Multiple_2019, Mikaelian_Normal_1982, Vartdal_Linear_2019}.     
  
  A related approximation often used in geophysical studies neglects the acceleration term in the fluid equation of motion~\cite{Lev_Rayleigh-Taylor_2008, Mondal_Propagator_2017} 
  %(henceforth referred to as the no-acceleration approximation) 
  and is also discussed here and compared with the exact numerical results. We show that this no-acceleration approximation is good when the physical length scale ($\lambda$) of the problem is much smaller $\lambda\ll(\nu^2/g)^{1/3}$ than the viscosity length-scale, as in many geophysical settings. On the other hand, the inviscid-flow approximation is good in the exact opposite viscosity limit. Therefore, the computational simplicity of the inviscid-flow approximation-based transfer-matrix formalism developed here is of potential use in physics.    
  
%%%%%%%%%%%%%%%%%%%%%%%%%%%%%%%%%%%%%%%%%%%%%%%%%%%%%%%%%%%
\paragraph*{Formalism.}

The equations that govern the instability dynamics of a viscous fluid in the incompressible limit~\cite{Chandrasekhar_Hydrodynamic_1961} are the linearized Navier-Stokes equation 
\begin{equation}\label{eq:flow}
	\rho\frac{\partial\vec{v}}{\partial t}=\nabla\cdot\sigma +\vec{f},
\end{equation}
the incompressible fluid constraint 
\begin{equation}\label{eq:incompressible}
	\nabla\cdot\vec{v}=0,
\end{equation} 
and the constitutive relations 
\begin{equation}\label{eq:constitutive}
	\sigma_{ij}=-p\delta_{ij}+
	\mu\left(\partial_iv_j+\partial_jv_i\right).
\end{equation}
The symbols $\sigma_{ij}, p,\vec{v},\vec{f}$ denote the deviation from the equilibrium state of the viscous stress tensor, fluid pressure, fluid velocity, and the external force density acting on the fluid. The density $\rho$ is the unperturbed static density of the fluid in equilibrium. We denote the perturbed density by $\delta\rho$, which satisfies the continuity equation. 
\begin{equation}\label{eq:continuity}
	\frac{\partial}{\partial t}\delta\rho=-\vec v\cdot\nabla\rho.
\end{equation} 
We choose the $z$-coordinate direction ($\hat z$) along the local gravitational field. With this choice the gravitational force density
\begin{align*}
	\vec f=\delta\rho\; g\;\hat z.
\end{align*} 

\begin{figure}[!]
 \includegraphics[scale=0.25]{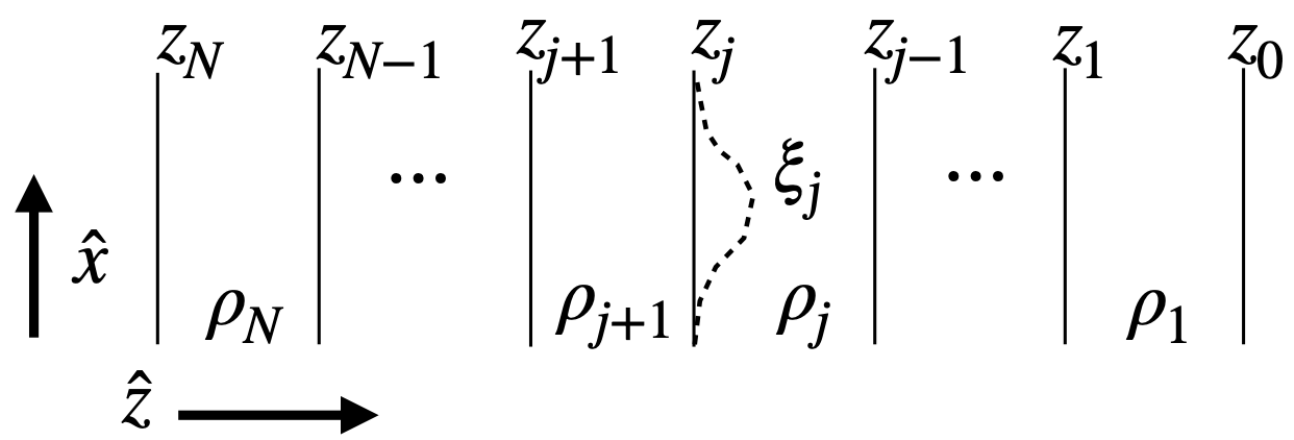}%
 \caption{\label{fig-0} Schematic of the multiple fluid layer system}
 \end{figure}

When density varies in the $\hat z$ direction, we may represent the variation by considering multiple layers, each with a constant density. We introduce the interface index $n=1,\dots,N$, so that the $n$-th layer is positioned between the interfaces $(z_{n+1},z_n)$, with $z_n>z_{n+1}$ (Fig.~\ref{fig-0}).
 Denoting the density of the layer $z_{j+1}\leq z \leq z_j$ by $\rho_{j+1}$ we can write the unperturbed density 
\begin{equation}\label{eq:density}
	\rho(z)=\hat z\sum_{j=0}^{N-1}\rho_{j+1}\Theta(z-z_{j+1})\Theta(z_j-z),
\end{equation}
where $\Theta$ denotes the Heaviside step function: $\Theta(x\geq0)=1$ and $\Theta(x<0)=0$. Eq.~(\ref{eq:density}) and (\ref{eq:continuity}), together with the relation between the perturbed velocity in the $\hat z$-direction and the interface displacement $\xi(z,x,t)$,
\begin{equation}\label{eq:xi-vel}
		\partial_t\xi(z,x,t)\equiv v_z(z,x,t),
\end{equation} 
determine the perturbed density
\begin{align}\label{eq:perturbed-density}
	\delta\rho(z,x,t)=\xi(z,x,t)\sum_{j=1}^{N-1}\Delta\rho_j\delta(z-z_{j}),
\end{align}
where $\Delta\rho_j\equiv(\rho_{j+1}-\rho_j)$ and $\delta(z)$ is the Dirac delta function. Our formalism also applies to a stratified heterogeneous fluid system, in which case the normal stresses acting on the two sides of an interface must differ by the {\em surface tension} $T^{(s)}/R_c$. In the linearized theory, the inverse interface radius of curvature $1/R_C\approx\partial_x^2\xi$, and the surface tension coefficient of the $j$-th interface is denoted as $T^{(s)}_j$. The force density $\vec f\equiv \hat z\;f_z$ in the presence of surface tension is:
\begin{align}\label{eq:fz}
	f_z=\sum_j\delta(z-z_j)(g\Delta\rho_j+T^{(s)}_j\partial_x^2)\xi
\end{align} 
 To linear order in perturbation theory, the velocity and interface fluctuation can be analyzed in terms of their normal modes, whose dependence on the planar coordinate and time is of the form
\begin{equation}\label{eq:normal-mode}
	h(z,x,t)\equiv \sum_k h_k(z)\;e^{\left(ikx+\gamma t\right)}.
\end{equation}
Here, $k$ denotes the interface-directed wave vector, $h_k$ denotes the $k$-th mode of the perturbations $\vec v, p, \sigma_{ij}, \xi$, and $\gamma$ is the growth rate of the $k$-th mode. Although we have chosen a 2D geometry for simplicity, our results apply to 3D Cartesian geometry by generalizing the 1D vector $k$ to a 2D vector. 

We introduce reference scales for viscosity ($\mu_0$), density ($\rho_0$), and length ($d_0$), using which we can define the time scale $\gamma_0^{-1}\equiv d_0^2/\nu_0$, where $\nu_0\equiv\mu_0/\rho_0$ is a kinematic viscosity scale. Using these in the Eq.~(\ref{eq:flow}-\ref{eq:constitutive}) we obtain the following relations between the normal mode amplitudes of the perturbed quantities: 

 \begin{alignat}{3}
 	\frac{d}{d z}\bm{\Psi}&=\bm{K}\cdot\bm{\Psi}-\bm{\Omega},~~{\rm where}\label{eq:pert-flow-a}\\
 	\bm{\Psi}&\equiv
 	\begin{pmatrix}
 	\tilde v_{z}\\ \tilde v_{x}\\ \frac{\tilde\sigma_{zz}}{2\mu_0 k}\\ \frac{\tilde\sigma_{xz}}{2\mu_0 k}
 	\end{pmatrix},\,\,  \bm{\Omega}\equiv\begin{pmatrix}0\\0\\ \tilde f_z/{2\mu_0 k}\\0\end{pmatrix}\nonumber\\
 	\bm{K}&\equiv\begin{pmatrix}0 & k & 0  & 0\\
 	-k & 0 & 0 & k\frac{2\mu_0}{\mu}\\
 	+\frac{\rho\gamma/\gamma_0}{2\rho_0 (kd_0)^2}k\ & 0 & 0 & k\\ 0 & k\left(\frac{2\mu}{\mu_0}+\frac{\rho\gamma/\gamma_0}{2\rho_0(k d_0)^2}\right)\ &\ -k & 0\end{pmatrix}.\nonumber 
 \end{alignat}
Here, $\tilde v_x\equiv v_{x,k}+v_{x,k}^*$, $\tilde v_z\equiv i(v_{z,k}^*-v_{z,k})$, $\tilde\sigma_{xz}\equiv \sigma_{xz,k}+\sigma_{xz,k}^*$, $\tilde\sigma_{zz}\equiv i(\sigma_{zz,k}^*-\sigma_{zz,k})$, and $\tilde f_z\equiv i(f_{z,k}^*-f_{z,k})$, are the real-valued components of the normal mode amplitudes. The vector $\bm{\Psi}$ spans both the velocity and stress sub-spaces and the vector $\bm{\Omega}$ is non-zero in the stress sub-space only. Note that the matrix $\bm{K}$ is $z$-dependent through the viscosity $\mu$ and the density $\rho$. Eq.~(\ref{eq:pert-flow-a}) has a formal solution
\begin{align*}
	\bm{\Psi}=P[e^{\int_{z_N}^z dz'\bm{K}(z')}]\bm{\Psi}_N
	- \int_{z_N}^z\!\! dz' P[e^{\int_{z'}^z dz''\bm{K}(z'')}] \bm{\Omega}(z'),
\end{align*}
where $P[e^{\int_{z_N}^z dz'\bm{K}(z')}]$ denotes path ordering of the power series expansion of the exponential, with matrices in each term ordered so that those evaluated at higher values of $z'$ stand to the left~\cite{peskin:1995}.  
\paragraph*{Transfer-Matrix approach for a Stratified heterogeneous fluid.}
For the case of a stratified fluid with density given by Eqn.~(\ref{eq:density}), we can rewrite the above equation: 
\begin{align}\label{eq:layered-soln}
\bm{\Psi}_j=P[\prod_{\ell=j+1}^{N}\!\! e^{\bm{K}_\ell d_\ell}]\bm{\Psi}_N\! -\!\!\sum_{m=j+1}^{N-1} P[\prod_{\ell=j+1}^{m>j}\!\! e^{\bm{K}_\ell d_\ell}]\bm{\Omega}_m,
\end{align} 
where $\bm{\Omega}_m\equiv\left(0,0,\tilde\xi_m\;\frac{g\Delta\rho_m-k^2T^{(s)}_m}{2\mu_0 k},0\right)^T$, the layer depth $d_\ell\equiv z_{\ell-1}-z_\ell$, $\bm{K}_\ell\equiv\frac{1}{d_\ell}\int_{z_\ell}^{z_{\ell-1}}dz\;\bm{K}(z)$, and $P[\prod_{\ell=m}^{n\geq m} e^{\bm{K}_\ell d_\ell}]=e^{\bm{K}_m d_m} e^{\bm{K}_{m+1} d_{m+1}}\dots e^{\bm{K}_n d_{n}}$. 

To find the growth rate dispersion $\gamma$ we need to relate $\tilde v_z$ component of $\Psi_j$ and $\tilde\xi_m$ component of $\Omega_m$ for all the layers. To accomplish this we introduce the projector matrices into the velocity and stress subspaces of $\bm{\Psi}$:
\begin{alignat}{2}
\hat P_v\equiv\begin{pmatrix}
	1 & 0 & 0 & 0\\
	0 & 1 & 0 & 0
\end{pmatrix};\,\, \hat P_\sigma\equiv \begin{pmatrix}
	0 & 0 & 1 & 0\\
	0 & 0 & 0 & 1
\end{pmatrix}	;\,\, \hat P_\sigma^T\equiv \begin{pmatrix}
	0 & 0\\ 0 & 0\\
	1 & 0 \\ 0 & 1
\end{pmatrix}\nonumber	
\end{alignat}
We use the boundary conditions that the topmost and bottom-most layers are immobile: $\vec v(z,x,t)=0$ for all $(x,t)$ at $z=z_N$ and $z=z_0$. As a result, both $\bm{\Psi}_N$ and $\bm{\Psi}_0$ can only be non-zero in the stress-subspace: $\bm{\Psi}_N=\hat P_\sigma^T\cdot\hat P_\sigma\bm{\Psi}_N$, and $\hat P_v\bm{\Psi}_0=0$. Using these relations in Eq.~(\ref{eq:layered-soln}) allows us to relate $\bm{\Psi}_N$ to $\bm{\Omega}_m$. Thus, we can simplify Eq.~(\ref{eq:layered-soln}) to give us the relation between the external stress and the velocity of the $j$-th layer:
\begin{align*}
	\hat P_v\bm{\Psi}_j&=
	\sum_{m=1}^{N-1}\Bigg(\bm{A}_{j+1}^N\left(\bm{A}_1^N\right)^{-1}\bm{A}_1^m%\nonumber\\
	 - \bm{A}_{j+1}^m\Bigg)\hat P_\sigma\bm{\Omega}_m,
\end{align*}
where $\bm{A}_{m}^n\equiv \hat P_v\cdot P[\prod_{\ell=m}^{n\geq m} e^{\bm{K}_\ell d_\ell}]\cdot\hat P_\sigma^T$. Note that $\bm{A}_{m}^n=0$, when $n<m$. Taking the dot product of the above equation with the vector $\hat e_1\equiv (1,0)$, and using the relations 
\begin{align}\label{eq:omegam}
	\hat P_\sigma\bm{\Omega}_m\equiv \hat e_1^T\cdot\tilde\xi_m\omega_m, ~~\omega_m\equiv\frac{\Delta\rho_m g-k^2 T^{(s)}_m}{2\mu_0 k},
\end{align}
and
\begin{align}\label{eq:tildevz}
	\tilde v_z\equiv\partial_t\tilde\xi=\gamma\tilde\xi,
\end{align}
we obtain the equation that determines the growth rate dispersion relation: 
\begin{alignat}{2}\label{eq:dispersion-det}
	&\mathrm{det}\left[\bm{G}-\gamma\right]=0\\
	&\bm{G}^m_j\equiv\hat e_1\cdot\Big(\bm{A}_{j+1}^N\left(\bm{A}_1^N\right)^{-1}\bm{A}_1^m 
	 - \bm{A}_{j+1}^m\Big)\cdot\hat e_1^T\;	 \omega_m\nonumber
\end{alignat}
Here the $(N-1)\times(N-1)$ matrix $\bm{G}$ has the row and column indices $j,m=1,\dots,N-1$. The matrix $\bm{G}$ depends on $\gamma$, so the problem is not analytically solvable even for the single interface (two-layer) problem. 

\begin{figure*}
 \includegraphics[scale=0.6]{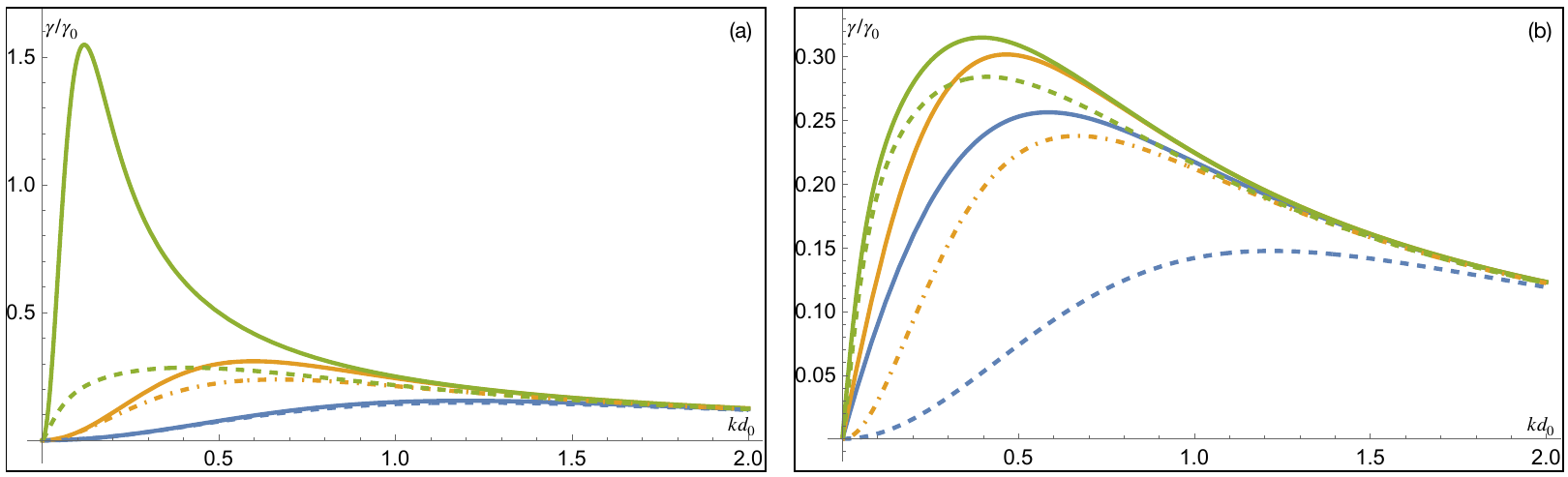}%
 \caption{\label{fig-1} Two-layer case for Atwood ratio $\frac{\rho_2-\rho_1}{\rho_1+\rho_2}=0.5$ and uniform kinematic viscosity $\nu/\nu_0=1$. We choose $gd_0/\nu_0=\gamma_0$. Dashed lines show exact results while the solid lines show (a) no-acceleration approximation, and (b) inviscid-flow approximation, for layer thickness $h=2d_0$ (blue, bottom-most set of dashed and solid curves), $h=4d_0$ (orange, middle set of dashed and solid curves), $h=20d_0$ (green, topmost set of dashed and solid curves).}
 \end{figure*}

We can use the general formalism to solve the two-layer case, $N=2$.
In Fig.~\ref{fig-1} we plot the numerical solution of Eq.~(\ref{eq:dispersion-det}) in the absence of surface tension, and compare it with two different approximations (see below) that yield analytic dispersion relations. 
\paragraph*{No-Acceleration Approximation.}
%%%%%%%%%
In the matrix $\bm{K}$ it is evident that when the effective kinematic viscosity $\nu\gg\gamma/k^2$, we can neglect the $\gamma$-dependence of the $\bm{K}$-matrix. The quantity $\gamma/\nu k^2\equiv Re$, the Reynolds number for instability dynamics. It is known that in the low Reynolds number ($Re\ll 1$) limit relevant, for example, in geophysical situations~\cite{Lev_Rayleigh-Taylor_2008}, the inertial term can be neglected. This amounts to dropping the fluid acceleration term in Eq.~(\ref{eq:flow}), which then becomes the Stokes equation  
\begin{equation*}\label{eq:no-flow}
	0=\nabla\cdot\sigma +\vec{f}.
\end{equation*}
%
%%%%
 An analytic expression for the transfer matrix elements of Eq.~(\ref{eq:dispersion-det}) can be obtained in the case of equal layer thickness $h$: 
\begin{align*}
	\mu_1\bm{A}^1_1=\mu_2\bm{A}^2_2=\begin{pmatrix}
		s_q-qc_q & -q s_q \\ q s_q & q c_q+s_q
	\end{pmatrix}
\end{align*}
\begin{align*}
	\bm{A}^2_1\!=\!\frac{\mu_1+\mu_2}{2\mu_1\mu_2}\!\begin{pmatrix}
		s_{2q}-2qc_{2q} & 2q\left(q\frac{\mu_2-\mu_1}{\mu1+\mu_2}-s_{2q}\right)\\
		2q\left(q\frac{\mu_2-\mu_1}{\mu1+\mu_2}+s_{2q}\right) & 2q c_{2q}+s_{2q}
	\end{pmatrix}.
\end{align*}
Here $q\equiv k h$, $c_q\equiv\cosh(q)$, $s_q\equiv\sinh(q)$.
 Using these in (\ref{eq:dispersion-det}) we find the growth rate dispersion: 
 \begin{align}\label{eq:aprx-dispersion}
 	\gamma &=\omega
 	\frac{\left((2\cosh (2 q)-\left(4 q^2+2\right)\right) (\sinh (2 q)-2 q)}{\left(\cosh (4 q)-8 q^2-1\right)+\frac{(\mu_1-\mu_2)^2}{\left(\mu _1+\mu _2\right)^2 }\left(8 q^4\right) },
 \end{align}
where $q\equiv kh$, and $\omega\equiv(\Delta\rho_1g-k^2T^{(s)}_1)/2k(\mu_1+\mu_2)$. In Fig.~\ref{fig-1}(a) we plot the above dispersion and compare it with the exact result Eq.~(\ref{eq:dispersion-det}). 
The two dispersions overlap over the entire range of wavevectors for small values of the scaled layer thickness $h/d_0$. Thus, the no-acceleration approximation allows us to obtain a closed-form expression for the two-layer dispersion of the Rayleigh-Taylor instability for a range of thicknesses. 
  
\begin{figure*}
 \includegraphics[scale=0.6]{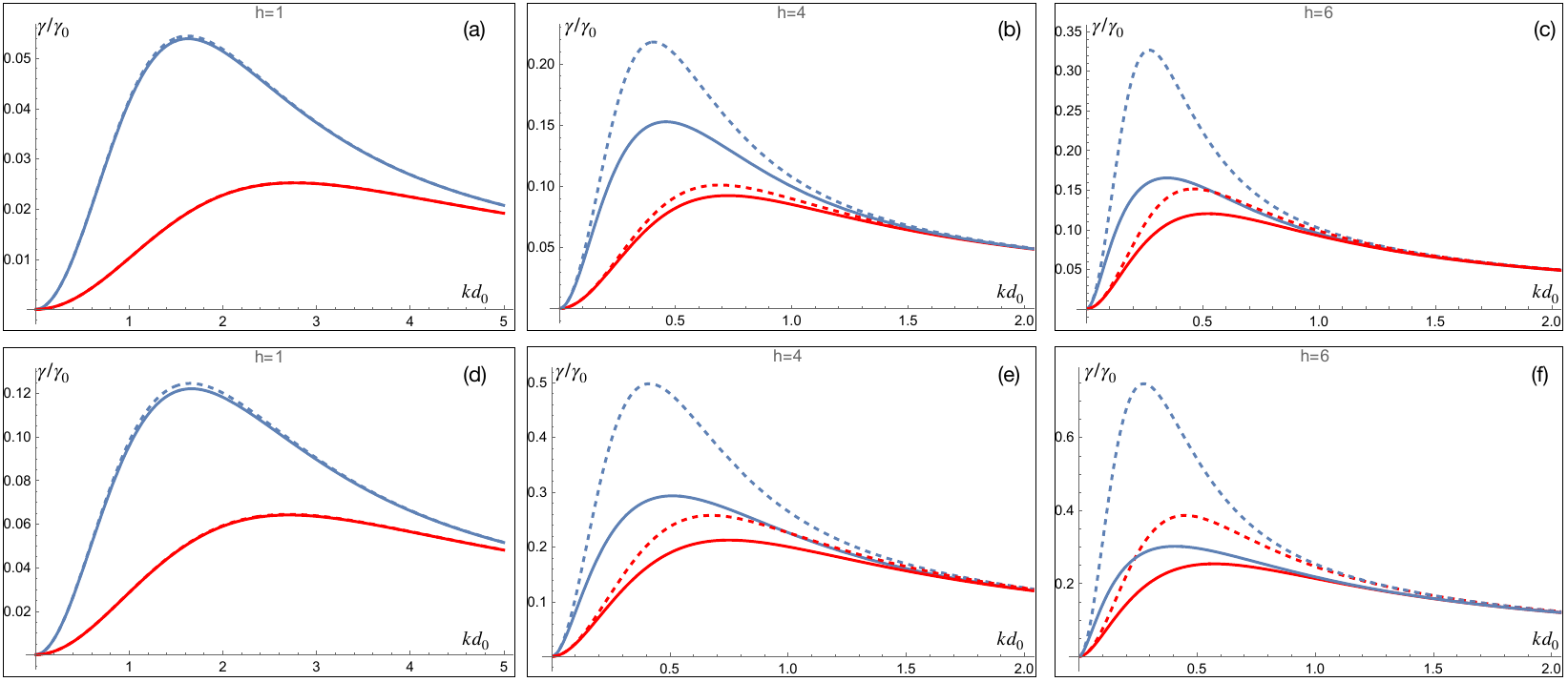}%
 \caption{\label{fig-2} The two dispersion branches in the three-layer case for the exact (solid lines) and the no-acceleration approximation (dashed lines). We choose $gd_0/\nu_0=\gamma_0$. The Atwood ratio across both interfaces is taken to be the same ($A_T=\frac{\rho_2-\rho_1}{\rho_1+\rho_2}=\frac{\rho_3-\rho_2}{\rho_2+\rho_3}$) and the kinematic viscosity is also uniform. For (a)-(c) $A_T=0.2$, and the layer thickness is $h=d_0$,$h=4d_0$, and $h=6d_0$, as indicated by the label h. For (d)-(f) $A_T=0.5$; h indicates the layer thickness.}
 \end{figure*}

  It is also evident from the plots that for large layer thickness $h$, the approximation gives a peak growth rate that increases with $h$ and deviates from the exact solution. This deviation is unsurprising since the no-acceleration approximation is good only when viscosity dominates. Indeed the $h\to\infty$ limit of Eq.~(\ref{eq:aprx-dispersion}) does not exist in the absence of viscosity, unlike in the exact case, Eq.~(51) of Ref.~\cite{Chandrasekhar_Hydrodynamic_1961}.    
%%%%
In the three-layer case Eq.~(\ref{eq:dispersion-det}) yields an expression for both branches of the dispersion relation. This solution is also plotted in Fig.~\ref{fig-2} and can be seen to deviate from the exact results for the three-layer case for large values of layer thickness.  
%%%%%%%%%%%%%%%
%
%%%%%%%%%%%%%%%%%%%%%%%%%%%%%%%%%%%%%%%%%%%%%%%%%%%%%%%%%%%%%%%%%%%%%%%%%%%%%%%%%%%
\paragraph*{Inviscid flow approximation.}
The inviscid flow approximation,
 \begin{equation}\label{eq:inviscid}
 	\nabla^2 \vec v=0,
 \end{equation}
is known to yield~\cite{Mikaelian_Rayleigh-Taylor_1996,Piriz_Rayleigh-Taylor_2006} the dispersion for the single interface problem with similar asymptotic behavior as the exact numerical solution. 
We show that the inviscid flow approximation considerably simplifies the transfer-matrix formalism. In the uniform kinematic viscosity limit, the problem of finding the dispersion reduces to a simple eigenvalue problem. We illustrate our approach by finding the growth rate for the case of three layers of fluid, each with a different viscosity and density. 
  Using equations~(\ref{eq:incompressible})-(\ref{eq:constitutive}), (\ref{eq:normal-mode}) and (\ref{eq:inviscid}) in the fluid equation of motion~(\ref{eq:flow}) we obtain the following first-order differential equation governing the evolution of the normal modes of the vertical fluid velocity and stress (assembled as the vector $\Phi$):
 \begin{alignat}{3}\label{flow-eqns}
 	\frac{d}{dz}\bm{\Phi}&=\bm{\Gamma}+k\;\bm{M}\cdot\bm{\Phi},~~{\rm where}\\
 	\bm{\Phi}&\equiv
 	\begin{pmatrix}
 	\tilde v_z\\ \frac{\tilde\sigma_{zz}}{2\mu_0 k}
 	\end{pmatrix},\,\,  \bm{\Gamma}\equiv\begin{pmatrix}0\\ \frac{\tilde f_z}{2\mu_0k}\end{pmatrix}\nonumber\\
 	\bm{M}&=\begin{pmatrix}0\ &\ \left(\frac{\mu}{\mu_0}+\frac{\rho}{2\rho_0}\frac{\gamma/\gamma_0}{k^2d_0^2}\right)^{-1}\\ \left(\frac{\mu}{\mu_0}+\frac{\rho}{2\rho_0}\frac{\gamma/\gamma_0}{k^2d_0^2}\right)\ &\ 0\end{pmatrix}.\nonumber 
 \end{alignat}
As before, we can write down a formal solution: 
\begin{align*}%\label{eq:formal-soln-inviscid}
	\bm{\Phi}&=P[e^{\int_{z_N}^z dz'k\bm{M}(z')}]\bm{\Phi}_N 
	-\!\!\int_{z_N}^z dz' P[e^{k\int_{z'}^z dz''\bm{M}(z'')}] \bm{\Gamma}(z').
\end{align*}
In the case of a stratified heterogeneous fluid, we find
\begin{align}
\bm{\Phi}_j=P[\prod_{\ell=j+1}^{N} e^{k\bm{M}_\ell d_\ell}]\bm{\Phi}_N - \sum_{m=j+1}^{N-1} P[\prod_{\ell=j+1}^{m>j} e^{k\bm{M}_\ell d_\ell}]\bm{\Gamma}_m,\nonumber
\end{align} 
where $\bm{\Gamma}_m\equiv\left(0,\tilde\xi_m\;\omega_m\right)^T$, the layer depth $d_\ell\equiv z_{\ell-1}-z_\ell$, and $\bm{M}_\ell\equiv\frac{1}{d_\ell}\int_{z_\ell}^{z_{\ell-1}}dz\;\bm{M}(z)$. 
Similar to Eq.~(\ref{eq:dispersion-det}) we find the equation that determines the dispersion in this approximation:
\begin{align}\label{eq:dispersion-det-inviscid-flow}
	\mathrm{det}&\Bigg[\Bigg({A}_{j+1}^N\left({A}_1^N\right)^{-1}{A}_1^m 
	- {A}_{j+1}^m\Bigg)
	\omega_m-\delta_j^m\gamma\Bigg]=0,
\end{align}
where ${A}_{m}^n\equiv\hat e_v\cdot P[\prod_{\ell=m}^{n\geq m} e^{k\bm{M}_\ell d_\ell}]\cdot\hat e_\sigma^T$, is a number, not an array, and the projectors $\hat e_{v,\sigma}$ into the one-dimensional velocity and stress sub-spaces are $\hat e_v=(1,0)$, and $\hat e_\sigma=(0,1)$. Introducing the parametrization
 \begin{align*}
 	e^{-\theta_\ell}\equiv\frac{\mu_\ell}{\mu_0}+\frac{\rho_\ell}{2\rho_0}\frac{\gamma/\gamma_0}{k^2d_0^2},
 \end{align*} 
 $\bm{M}_\ell\equiv\cosh\theta_\ell\;\bm{\sigma}_x + i\sinh\theta_\ell\;\bm{\sigma}_y$, where $\bm{\sigma}_{x,y}$ are Pauli matrices. Further, since $\bm{M}^2=\bm{1}$, we have the identity $e^{q\bm{M}}\equiv\cosh q\;\bm{1}+\sinh q\;\bm{M}$. The advantages of the inviscid flow approximation are obvious in that we have a $2\times 2$ matrix to deal with, rather than a $4\times 4$ matrix obtained in Ref.~\cite{Chandrasekhar_Hydrodynamic_1961}. This allows us to derive analytical results for cases with multiple fluid density and viscosity discontinuities. In particular, using $\hat e_v\cdot\sigma_{x}\cdot\hat e_\sigma^T=1=\hat e_v\cdot i\sigma_{y}\cdot\hat e_\sigma^T$, and $\hat e_v\cdot\sigma_{z}\cdot\hat e_\sigma^T=0=\hat e_v\cdot\bm{1}\cdot\hat e_\sigma^T$, the transfer matrix
\begin{align}
A_{m}^{n\geq m}&=\hat e_v
\cdot\prod_{\ell=0}^{n-m}\begin{pmatrix}
	c_{q_{m+\ell}} & s_{q_{m+\ell}} e^{\theta_{m+\ell}}\\
	s_{q_{m+\ell}} e^{-\theta_{m+\ell}} & c_{q_{m+\ell}}
\end{pmatrix} \cdot\hat e_\sigma^T
\end{align}
Here $q_j\equiv k d_j$, $c_{q_j}\equiv\cosh(q_j)$, and $s_{q_j}\equiv\sinh(q_j)$.
Thus, we find the general form of the transfer matrix elements relevant for the two- and three-layer cases: 
\begin{align}\label{eq:transfer-matrix-j}
A_j^j&=s_{q_j}e^{\theta_j}\,;\,
	A_j^{j+1}=s_{q_j}c_{q_{j+1}}e^{\theta_j}+s_{q_{j+1}}c_{q_{j}}e^{\theta_{j+1}}\\
	A_j^{j+2}&=\prod_{\ell=0}^2 s_{q_{j+\ell}}e^{(-1)^{\ell}\theta_{j+\ell}}
	 +\sum_{\sigma(012)}c_{q_j}c_{q_{j+1}}s_{q_{j+2}}e^{\theta_{j+2}},\nonumber
\end{align}
 where $\sigma(012)$ denotes cyclic permutation of the indices $j,j+1,j+2$.
  
For the two-layer case Eq.~(\ref{eq:dispersion-det-inviscid-flow}) becomes
\begin{align}\label{eq:disp-inviscid-2layer}
	{A}_{2}^2\left({A}_1^2\right)^{-1}{A}_1^1\;\frac{\Delta\rho_1g-k^2T^{(s)}_1}{2\mu_0 k}-\gamma=0, 
\end{align}
where the explicit form of ${A}_1^2$ and ${A}_1^1$  is given in Eq.~(\ref{eq:transfer-matrix-j}). 
We thus obtain the dispersion equation:
\begin{align*}%\label{eq:quadratic-gamma-2layer}
	\gamma^2 + 2k^2\gamma\;\frac{c_{q_1}\mu_1+c_{q_2}\mu_2}{c_{q_1}\rho_1+c_{q_2}\rho_2}
	-\frac{kg(\rho_2-\rho_1)-k^3T^{(s)}_1}{c_{q_1}\rho_1+c_{q_2}\rho_2}=0,
\end{align*}
which agrees with Eq.(25) of Ref.~\cite{Mikaelian_Rayleigh-Taylor_1996}. This dispersion is plotted and compared with the exact numerical solution in the absence of surface tension in Fig.~\ref{fig-1}(b). 
%%%
\paragraph*{Three-Layer case: Analytic results in the uniform kinematic viscosity limit.}
%%%
The three-layer case ($N=3$) involves the $2\times2$ matrix ${G}$:
\begin{align}\label{eq:mat-G}
	{G}_j^m&\equiv\Big({A}_{j+1}^3({A}_1^3)^{-1}{A}_1^m -
	 {A}_{j+1}^m\Big)\;\omega_m,
\end{align} 
where the interface indices $m,j=1,2$.
In the limit of uniform kinematic viscosity ($\mu_j/\rho_j\equiv\nu$)
 \begin{align*}
 	e^{-\theta_j}=\frac{\rho_j}{\rho_0}\left(\frac{\nu}{\nu_0}+\frac{\gamma}{\gamma_0}\frac{1}{2(kd_0)^2}\right)\equiv\frac{\rho_j}{\rho_0}\frac{\gamma^*}{\gamma_0},
 \end{align*}
using which we can factor out the $\gamma$-dependence from the transfer-matrix elements:  
\begin{align*}
	A_j^{j+1}\equiv\frac{\gamma_0}{\gamma^*}c_j^{j+1},\,
	A_j^{j}\equiv\frac{\gamma_0}{\gamma^*}c_j^j,\,
	A_1^3\equiv\frac{\gamma_0}{\gamma^*}c_1^3.
\end{align*}
The $\gamma$-dependence also factors out from the matrix ${G}$ in Eq.~(\ref{eq:mat-G}) and we can rewrite Eq.~(\ref{eq:dispersion-det-inviscid-flow}) to obtain the dispersion equation: 
\begin{align}\label{eq:mat-C}
\mathrm{det}&\Big(\frac{\gamma\gamma^*}{\gamma_0^2}-\bm{C}\Big)=0,\\
\bm{C}&=\begin{pmatrix}
	\frac{c_2^3c_1^1}{c_1^3}\frac{\Delta\rho_1}{2\rho_0kd_0} & \frac{c_1^1c_3^3}{c_1^3}\frac{\Delta\rho_1}{2\rho_0kd_0}\\ 	\frac{c_2^3c_1^2-c_2^2c_1^3}{c_1^3}\frac{\Delta\rho_2}{2\rho_0kd_0} & \frac{c_3^3c_1^2}{c_1^3}\frac{\Delta\rho_2}{2\rho_0kd_0}	
\end{pmatrix}\nonumber\\
c_1^1c_3^3&=c_2^3c_1^2-c_2^2c_1^3\equiv\frac{\rho_0}{\rho_1}\frac{\rho_0}{\rho_3}s_{q_1}s_{q_3}\nonumber\\
c_1^2c_3^3&\equiv\frac{\rho_0}{\rho_1}\frac{\rho_0}{\rho_3}c_{q_2}s_{q_1}s_{q_3}\Big[1+\frac{\rho_1}{\rho_2}\frac{\tanh(q_2)}{\tanh(q_1)}\Big]\nonumber\\
c_2^3c_1^1&\equiv\frac{\rho_0}{\rho_2}\frac{\rho_0}{\rho_1}c_{q_3}s_{q_2}s_{q_1}\Big[1+\frac{\rho_2}{\rho_3}\frac{\tanh(q_3)}{\tanh(q_2)}\Big]\nonumber
\end{align}
The matrix $\bm{C}=(\gamma^*/\gamma_0){G}$ is independent of $\gamma$ and Eq.~(\ref{eq:mat-C}) determines the dispersion simply if we diagonalize the matrix $\bm{C}$. Denoting the diagonal entries of $\bm{\Sigma}$ as $\Sigma_d$, the  dispersion relation is a simple quadratic equation: 
\begin{align}\label{eq:disp-3layer}
	\left(\frac{\gamma}{\gamma_0}\right)^2\frac{1}{2(kd_0)^2} +\frac{\gamma}{\gamma_0}\frac{\nu}{\nu_0} -\Sigma_d=0. 
\end{align}
 \begin{figure*}[!]
 \includegraphics[scale=0.6]{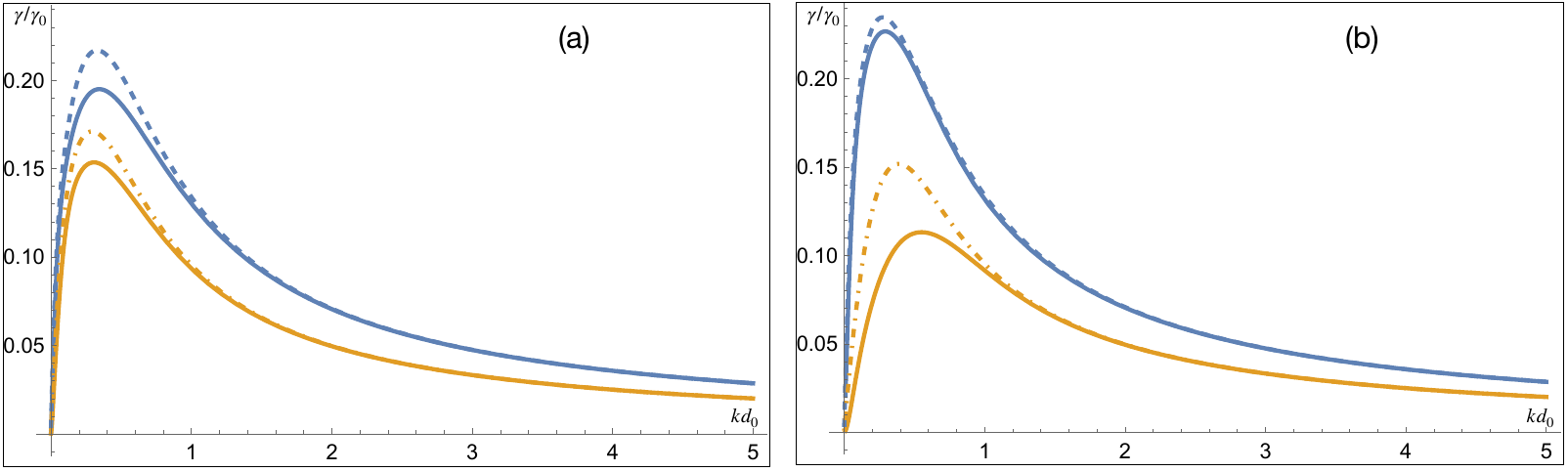}%
 \caption{\label{fig-3} The two branches of dispersion $\gamma/\gamma_0$ {\em vs} $kd_0$ for three-layers. Dashed lines are the inviscid-flow approximation and solid lines are exact results. We choose $gd_0/\nu_0=\gamma_0$. The Atwood ratio across the top interface is $\frac{\rho_3-\rho_2}{\rho_3+\rho_2}=0.2$, and that across the bottom interface is $\frac{\rho_2-\rho_1}{\rho_1+\rho_2}=0.5$. The uniform kinematic viscosity $\nu/\nu_0=1$. For (a) all the layers have the same thickness $h=20d_0$. For (b) the middle-layer thickness is reduced to $h=4d_0$.}
 \end{figure*}

Thus, each (positive/negative) eigenvalue of $\bm{C}$ gives a (positive/negative) branch of the dispersion. When all layers have the same thickness $h$ we find the two eigenvalues of $\bm{C}$: 
\begin{align*}
	\Sigma_\pm&=\frac{\Bigg[1\pm \sqrt{1+\frac{\Delta\rho_1\Delta\rho_2}{(\rho_3-\rho_1)^2}[\frac{1}{\cosh^{2}(kh)}-\frac{(\rho_1+\rho_2)(\rho_2+\rho_3)}{\rho_2^2}]}\;\Bigg]}{2kd_0\left[\tanh(kh)\frac{\rho_2}{(\rho_3-\rho_1)}+(\frac{\rho_1+\rho_3+\rho_1\rho_3/\rho_2}{(\rho_3-\rho_1)})\coth(kh)\right]}.
\end{align*} 

These results can be generalized to include surface tension by substituting $\Delta\rho_{1,2}\to\Delta\rho_{1,2}-k^2T^{(s)}_{1,2}$, and $\rho_3-\rho_1\to \rho_3-\rho_1-k^2(T^{(s)}_1+T^{(s)}_2)$.
Note that when $\rho_1=\rho_2$ (and $T^{(s)}_1=0$), or $\rho_3=\rho_2$ (and $T^{(s)}_2=0$), the system is identical to a two-layer system and the eigenvalue $\Sigma_-\to 0$. From Eq.~(\ref{eq:disp-3layer}) we see that when $\nu\to0$, and $h\to\infty$ the growth rate $\gamma^2=\gamma_0^2 kd_0 A_T$ -- identical to Eq.~(51) of Ref.~\cite{Chandrasekhar_Hydrodynamic_1961} for $\gamma_0^2d_0\equiv g$. 

In Fig.~\ref{fig-3} we plot the dispersion for different Atwood ratios for equal layer thickness and uniform viscosity. As can be seen from Eq.~(\ref{eq:disp-3layer}), when the Atwood ratios across the two interfaces are different the two branches remain separated even at short wavelengths. Further, when the middle layer thickness is reduced, the branches are ``repelled" away from each other. In the limit of a thin middle layer, one of the branches vanishes, as discussed below. In all cases, the inviscid-flow approximation overestimates the peak growth rate, with the error increasing with decreasing layer thickness.   
\paragraph*{Analytic results in the zero viscosity limit:}
In the limit of zero viscosity the parameters 
 \begin{align*}
 	e^{-\theta_j}=\frac{\rho_j}{\rho_0}\frac{\gamma}{\gamma_0}\frac{1}{2(kd_0)^2},
 \end{align*}
and we can rewrite Eq.~(\ref{eq:dispersion-det-inviscid-flow}) to obtain the following equation for determining the dispersion:
\begin{align}\label{eq:mat-C0}
	\mathrm{det}&\Big(\gamma^2-2k^2d_0g\;\bm{C}\Big)=0.
\end{align}
We now consider the case, when the middle-layer is a thin region in which the density smoothly interpolates between the top-layer density $\rho_3$ and the bottom-layer density $\rho_1$. Taking the smooth interpolation to be given by the exponential function $\rho(z)=\rho_3\;e^{(z_2-z)/\lambda_m}$, and the average density in the thin middle-layer $\rho_2=\frac{1}{d_2}\int_{z_2}^{z_1}dz\;\rho(z)$, we find the interpolation length scale $\lambda_m\equiv\rho_2 d_2/\Delta\rho$, where $\Delta\rho\equiv\rho_3-\rho_1$. As long as $d_2\leq 2\lambda_m$ (or $\Delta\rho\leq 2\rho_2$) we also have $\rho_2\approx\sqrt{\rho_3\rho_1}$. In this thin middle-layer ($kd_2\ll kd_{1,3}$) limit, we find
 \begin{align*}
 	2k^2d_0 g\;\bm{C}=\frac{gk}{1+kd_2\left(\frac{\rho_1\rho_3+\rho_2^2}{\rho_2(\rho_1+\rho_3)}\right)}\begin{pmatrix}
 		\frac{\Delta\rho_1}{\rho_1+\rho_3}\ &\ \frac{\Delta\rho_1}{\rho_1+\rho_3} \\
 		\frac{\Delta\rho_2}{\rho_1+\rho_3} & \frac{\Delta\rho_2}{\rho_1+\rho_3} 
 	\end{pmatrix}
 \end{align*}
 Diagonalizing $\bm{C}$ and using the above-mentioned relations we find the growth rate from Eq.~(\ref{eq:mat-C0}):
 \begin{align}\label{eq:gamma-thin-layer}
 	\gamma^2=\frac{gkA_T}{1+2A_Tk\lambda_m}
 \end{align} 
 This is identical to Eq.(3) of Ref~\cite{Betti_Growth_1998} when we identify $2\lambda_m$ as the minimum density gradient length scale. Thus, we see that the long-wavelength ($k\lambda_m\ll 1$) modes are identical to the two-layer case in the zero viscosity limit, while the short-wavelength modes ($k\lambda_m\gg 1$) are non-dispersive. 

The procedure for finding the dispersion relation in the limit of uniform kinematic viscosity can be generalized to $N$-layers. In the case of uniform kinematic viscosity, the problem reduces to a standard eigenvalue problem: Eq.~(\ref{eq:mat-C}) determines all the $2(N-1)$ branches of the dispersion when we diagonalize the $(N-1)\times(N-1)$ dimensional matrix $\bm{C}=(\gamma^*/\gamma_0){G}$.   
%%%%%%%%%%%%%%%%%%%%%%%%%%%%%%%%%%%%%%%%%%%%%%%%%%%%%%
\paragraph*{Summary:}
%%%%%
In summary, we have presented a transfer-matrix formalism for solving the linearized perturbation theory of the Rayleigh-Taylor instability. The formalism applies to the case of spatially varying density and viscosity, which is a more natural scenario than the classical limit of a single interfacial density jump. The theory has the advantage of being computationally straightforward. It generalizes the physical insights gained from the study of the single-interface problem and is particularly useful in the limits of zero viscosity and uniform kinematic viscosity. In both these limits, we find analytic results for the growth of the two-interface Rayleigh-Taylor instability. We use these results to derive the growth rate for the case of two interfaces and show that as the interfaces come closer together, the two dispersion branches are ``repelled", one branch growing at the expense of the other. In the limit of a thin middle layer across which the density smoothly interpolates between the top and bottom layer density, we recover the known asymptotic dispersion of the single dispersive branch. The formalism developed here has potential applications for studying the effect of spatially varying density profiles on the growth of the Rayleigh-Taylor instability.
\bibliography{RTInstability}
%%%%%%%%%%%%%%%%%%%%%%%%%%%%%%%%%%%%%%%%%%%%%%%%%%%%%%%%%%%%%%%%%%%%%%

%%%%%%%%%%
\end{document}